\documentclass[fleqn,10pt]{wlscirep}
\pdfoutput=1                % to force PDFLaTeX %

\usepackage{amsmath}        % need ams stuff    %
\usepackage{amsfonts}       % to make equations %
\usepackage{amssymb}        % look nice         %
\usepackage{upgreek}
\usepackage{listings}
\usepackage{hyperref}
\usepackage{caption}
\title{VOEvent Standard for Fast Radio Bursts}
\author[*,1]{Emily Petroff}
\author[$\dagger$,2,3]{Leon Houben}
\author[4]{Keith Bannister}
\author[5,6]{Sarah Burke-Spolaor}
\author[7]{Jim Cordes}
\author[3,1,2]{Heino Falcke}
\author[8]{Ronald van Haren}
\author[9,10,11]{Aris Karastergiou}
\author[2,12]{Michael Kramer}
\author[13]{Casey Law}
\author[1,14]{Joeri van Leeuwen}
\author[5,6]{Duncan Lorimer}
\author[8]{Oscar Martinez-Rubi}
\author[3]{J\"org Rachen}
\author[2]{Laura Spitler}
\author[15]{Amanda Weltman}

\affil[1]{ASTRON, the Netherlands Institute for Radio Astronomy, Oude Hoogevensedijk 4, 7991 PD Dwingeloo, The Netherlands}
\affil[2]{Max-Planck-Institut f\"ur Radioastronomie, Auf dem H\"ugel 69, D-53121 Bonn, Germany}
\affil[3]{Department of Astrophysics/IMAPP, Radboud University, PO Box 9010, 6500 GL Nijmegen, The Netherlands}
\affil[4]{Australia Telescope National Facility, CSIRO Astronomy and Space Science, PO Box 76, Epping, NSW 1710, Australia}
\affil[5]{Department of Physics and Astronomy, West Virginia University, PO Box 6315, Morgantown, WV 26506, USA}
\affil[6]{Center for Gravitational Waves and Cosmology, West Virginia University, Chestnut Ridge Research Building, Morgantown, WV 26505, USA}
\affil[7]{Astronomy Department, Cornell University, Ithaca, NY 14853}
\affil[8]{Netherlands eScience Center, Science Park 140 (Matrix I), 1098 XG Amsterdam, The Netherlands}
\affil[9]{Astrophysics, University of Oxford, Denys Wilkinson Building, Keble Road, Oxford OX1 3RH, UK}
\affil[10]{Department of Physics and Electronics, Rhodes University, PO Box 94, Grahamstown 6140, South Africa}
\affil[11]{Physics Department, University of the Western Cape, Cape Town 7535, South Africa}
\affil[12]{Jodrell Bank Centre for Astrophysics, University of Manchester, Alan Turing Building, Oxford Road, Manchester M13 9PL, United Kingdom}
\affil[13]{Department of Astronomy and Radio Astronomy Lab, University of California, Berkeley, CA, USA}
\affil[14]{Anton Pannekoek Institute, University of Amsterdam, PO Box 94249, 1090 GE Amsterdam, The Netherlands}
\affil[15]{Department of Mathematics and Applied Mathematics, University of Cape Town, South Africa}
\affil[*]{email: ebpetroff@gmail.com}
\affil[$\dagger$]{email: l.houben@astro.ru.nl}

\begin{abstract}
Fast radio bursts are a new class of transient radio phenomena currently detected as millisecond radio pulses with very high dispersion measures. As new radio surveys begin searching for FRBs a large population is expected to be detected in real-time, triggering a range of multi-wavelength and multi-messenger telescopes to search for repeating bursts and/or associated emission. Here we propose a method for disseminating FRB triggers using Virtual Observatory Events (VOEvents). This format was developed and is used successfully for transient alerts across the electromagnetic spectrum and for multi-messenger signals such as gravitational waves. In this paper we outline a proposed VOEvent standard for FRBs that includes the essential parameters of the event and where these parameters should be specified within the structure of the event. An additional advantage to the use of VOEvents for FRBs is that the events can automatically be ingested into the FRB Catalogue (FRBCAT) enabling real-time updates for public use. We welcome feedback from the community on the proposed standard outlined below and encourage those interested to join the nascent working group forming around this topic.
\end{abstract}

\begin{document}

\flushbottom
\maketitle
\thispagestyle{empty}

\section*{Introduction}
Fast radio bursts (FRBs) are one of the most exciting topics in modern astrophysics. FRBs are detected as millisecond radio pulses with a high dispersion measure (DM) defined as the frequency-dependent delay of the pulse arrival time across an observing bandwidth. Physically, this DM is related to the electron column density $n_e$ along the line of sight as
\begin{equation}
\mathrm{DM} = \int_{0}^{D} n_e d\ell 
\end{equation}
\noindent where D is the distance between the source and the observer along some path $\ell$. All known FRBs have DMs in excess of the modelled electron density contribution from the Milky Way, and all but one~\cite{Keane12} have DMs $>1.5 \times \mathrm{DM}_\mathrm{MW_{NE2001}}$ where $\mathrm{DM}_\mathrm{MW_{NE2001}}$ is the electron density contribution along the line of sight modelled by NE2001\cite{Cordes02}. This excess in DM suggests FRBs originate extragalactically leading to energetic progenitor theories such as binary neutron star mergers~\cite{Zhang2014}, collapses of neutron stars to black holes~\cite{Falcke}, extremely active young pulsars in nearby galaxies~\cite{Cordes2015}, and young magnetars in dense supernova remnants~\cite{Metzger2017}, to name a few. The designation of a bright single pulse as an FRB (as opposed to a bright single pulse from a Galactic pulsar) has been based on its DM.

The first FRB was discovered in 2007 by Lorimer et al.\cite{Lorimer07}, FRB 010724\footnote[9]{FRBs currently follow the date-based naming conventions for gamma-ray burst and gravitational wave events: FRB YYMMDD.}, and since then progress has increased rapidly. Twenty five FRB sources~\cite{frbcat} have been published\footnote[8]{All publicly available FRBs are included in the FRB Catalogue (FRBCAT); \url{http://www.frbcat.org}} and one source, FRB 121102, has been seen to repeat~\cite{Spitler2016}. Interferometric observations of pulses from FRB 121102 provided positional accuracy capable of localizing the burst and pinpointing the host galaxy -- a dwarf galaxy at a distance of $\sim$1 Gpc~\cite{Chatterjee2017}. This detection confirms an extragalactic progenitor for this burst and a highly energetic production process for the radio pulses. The progenitors of FRBs, however, remain unknown. A repeating burst rules out a cataclysmic progenitor for FRB 121102; however, this source may not be representative of the full population of FRBs as it is the only FRB that has been seen to repeat~\cite{P871}. Ultimately more FRBs need to be found, localized, and monitored to determine whether this observed behaviour is common. 

Recently, fast radio bursts have also been detected in real-time and followed-up with telescopes over radio, optical, X-ray, and $\upgamma$-ray wavelengths and one searched for multi-messenger signals from neutrinos~\cite{PetroffFRB,Keane2016,131104Radio,FRB150215}. Initial results have been inconclusive as the burst location within a large primary beam is unknown and association between the FRB and a multi-wavelength counterpart depends on detecting temporally coincident, distinctive transient emission. Some progenitor theories predict associated optical or X-ray transients and the search for possible counterparts remains essential. 

As more telescopes begin to detect FRBs in real-time, a standard way to circulate new detections becomes necessary to enable efficient observing by telescopes wishing to follow-up FRBs. A structure for the dissemination of astronomical transients called a VOEvent has been developed by the International Virtual Observatory Alliance (IVOA) which has been used to great effect for gamma-ray and supernova astronomy. In this paper we describe the structure of a VOEvent that may be used to describe FRBs, including some information on the contents of the FRB VOEvent and practical uses. In the following sections we present some motivations for the development of a VOEvent standard, describe the VOEvent framework and why it is well-suited to FRB detections, describe the particulars of the FRB event structure, elaborate on their different uses and explain the automatic ingestion of FRB VOEvents into the FRB Catalogue~\cite{frbcat}.

\section*{Motivation}

Many properties of FRBs remain unknown; a much larger population, with consistent monitoring and multi-wavelength follow-up, will be needed to answer questions related to their origins and possible associated emission. A rapid increase in the FRB detection rate is expected in the next few years as more telescopes begin searching for FRBs and as new wide-field interferometers come on-line. Next generation interferometric radio telescopes such as the Aperture Tile In Focus (Apertif) upgrade to the Westerbork Telescope in the Netherlands, the Canadian Hydrogen Intensity Mapping Experiment (CHIME) in North America, the Hydrogen Intensity and Real-time Analysis eXperiment (HIRAX) in South Africa, the Upgrade to the Molonglo Synthesis Telescope (UTMOST) and the Australian Square Kilometre Array Pathfinder \cite{ASKAPFRB} (ASKAP) in Australia are expected to detect hundreds to thousands of FRBs per year when operating at full sensitivity. Most FRB-finding experiments, such as the VLA realfast project\footnote{\url{http://realfast.io}}, also plan to detect these bursts in real-time, issuing triggers for follow-up upon robust detections.

At the time of writing only a handful of FRBs have been detected in real-time, all from the Parkes telescope, and only four have reported multi-wavelength follow-up. The current triggering procedure with detections at the Parkes telescope is to send an email with the relevant burst parameters to collaborators who then carry out observations. This process introduces considerable lag time between the discovery of an FRB and the first on-source observations with other telescopes. Efforts underway using the Effelsberg telescope and the Low Frequency Array (LOFAR) aim to capture the same burst in the bands of both telescopes. The large difference in observing frequency and the high DMs of FRBs allow for this detection, ensuring that the dispersed tail of a burst found with Effelsberg arrives seconds to minutes later in the LOFAR band. Therefore the ability to send triggers between the two telescopes on timescales of a few tenths of seconds is essential for this type of project to succeed. Currently, a simple VOEvent network links the two telescopes, sending triggers between the observatories. More information about the project is given in the first use case below.

%Added a little extra text here
A faster and more standardized method of alerting the community to new events must be implemented for FRBs in order to allow open and versatile reactions to FRB detections. Recent experiments have already shown that multi-wavelength follow-up may yield interesting results~\cite{ARTEMIS}, such as the detection of variable radio sources in the fields of FRB 150418 and FRB 131104~\cite{Keane2016,131104Radio} and of a 380-s $\upgamma$-ray transient temporally coincident with FRB 131104 in the field of the burst by the \textit{Swift} telescope~\cite{DeLaunay2016}. However, no definitive association has yet been made between an FRB and a multi-wavelength transient counterpart. Multi-wavelength emission on the shortest timescales post-burst is poorly constrained due to the high latency of current FRB alert methods. Machine issued, and machine parsed, event notifications would enable follow-up at these previously unexplored timescales. Additionally, recent observations of FRB 121102 have shown that at least this one FRB source goes through phases of high activity when repeat pulses are more likely to be detected. For such a source a detection alert can be followed with extensive monitoring for repeated bright bursts by other radio telescopes.
\newpage

The VOEvent structure provides a robust framework for sending and receiving event alerts upon the detection of a new FRB which can be used to trigger follow-up. VOEvents are already widely issued upon detection of gamma-ray bursts (GRBs) and supernovae and are beginning to be used to describe new gravitational wave detections with the Laser Interferometer Gravitational-Wave Observatory (LIGO). VOEvents have many advantages over other trigger dissemination methods. Firstly, VOEvents are embedded within the larger Virtual Observatory framework, discussed in more detail in the next section. Tools for distributing and receiving VOEvents, as well as code for parsing individual event files, are freely available on-line and well documented. Secondly, a standardized event structure for FRBs used by radio telescopes searching for new bursts gives greater flexibility to any multi-wavelength or multi-messenger facility wishing to trigger upon FRB detections. Thirdly, FRB VOEvents will be automatically added to the FRB Catalogue (FRBCAT) either as new entries upon detection or as updates to previously published bursts when additional detections are made. Details of this functionality are given in a separate section below. Our intention for this paper is to provide the necessary background on the VOEvent structure and establish a template for FRB event triggers.

\section*{Use case examples}
Here we provide three of many potential use cases where such a system would be useful to facilitate connections between multiple observatories.
%Case of triggering for lower frequency detection
\subsection*{Use case 1: Triggering LOFAR on Effelsberg detection}

The Effelsberg 100m telescope is currently being used to trigger the LOFAR transient buffer boards~\cite{LOFAR,Schellart2013} (TBBs) on potential FRB pulses detected at 1.4 GHz, which can enable an \textit{a posteriori} localisation of the FRB with arcminute resolution, if the pulses emit sufficiently strong signals in the LOFAR HBA frequency range (110--190 MHz). For a fiducial DM = 500, the time delay for the pulse to reach the center of the HBA band is about 90\,s, with about 90\,s tolerance to catch it somewhere in the frequency band. Thus, within roughly a minute, the pulse must be identified at the Effelsberg site and information about precise timing and a DM estimate, accurate to at least 50\%, must be extracted. Then, a VOEvent message is sent to the LOFAR system, which is able to receive the event, react to the message within about a second, and stop the TBBs at the optimal time. After this, a decision has to be taken whether to dump the TBB data, which involves a data volume of about 5 TBs for all LOFAR stations and takes about 30 minutes to complete. Here, a VOEvent \textit{update} message, providing information on RFI cleaning, improved DM estimate or other essential information like pulse broadening, can be used to estimate the chances of finding a usable signal in the data, and decide to either dump the data, or to restart the TBBs to wait for the next trigger.

%Case of triggering for optical follow-up
\subsection*{Use case 2: Robotic optical follow-up of Apertif detection}

Future searches with the Apertif system on the Westerbork telescope will search for FRBs over a 9 deg$^2$ field of view with an approximate error ellipse on any FRB detection with dimensions $25'' \times \leq 30'$. A single pulse search operating on the incoming data at the telescope will search for FRBs in real-time and provide a score for FRB candidates based on machine learning algorithms incorporated into the search. Upon detection of a highly robust candidate a VOEvent will be issued with the basic information about the FRB and an importance score corresponding to the significance of the candidate in the pipeline. In this case, a VOEvent will be issued within seconds of a detection and broadcast over the VOEvent Network where it may be received by subscribers. As an example, a robotic optical telescope may receive this event and decide to trigger based on time of day, estimated redshift, and FRB peak flux density and begin immediate follow-up observations to search for optical transients with high cadence imaging of the FRB field.

%Case for coordinated follow-up using a future FRB monitor
\subsection*{Use case 3: Coordinated follow-up campaigns and outreach}

Once FRB VOEvents are more commonly used, observatories will send around regular events not only upon a detection but also when they begin dedicated FRB search observations. In this capacity FRB VOEvents will enable more coordinated follow-up campaigns between observatories. A platform can be designed to collect, store, and visualise all messages sent about planned and ongoing FRB observations for observers to decide if and what to observe or whether to shadow a telescope searching for FRBs. Such coordination would increase the chance of detecting an FRB in multiple parts of the electromagnetic spectrum and might result in a more homogeneous distribution of the amount of follow-up hours per known FRB position. One can think about a system similar to the Radboud Radiolab's VLBI Monitor\footnote{\url{https://bitbucket.org/vlbi/} and \url{https://vlbimon1.science.ru.nl}} that uses JSON messages for coordinating large VLBI campaigns. Observers would be able to see the status of every collaborating observatory and swiftly respond to any changes with the actively participating telescopes. Such a platform could also be used for outreach purposes, providing a public visual aid to what the observer's telescope is doing (what part of the sky is being observed, for instance) or by involving amateur astronomers to perform shadowing observations with their optical or radio telescopes.

\section*{The VOEvent Framework}
The IVOA is an organisation with the mission to provide the international astronomical community with the organizational structures necessary to make data, tools, and communication about these assets available for everyone within the community. The goal is to make astronomical data so transparent that it appears to be the product of one virtual observatory (VO). Recommendations and standards are therefore provided to make (meta)data formats more uniform throughout the astronomical community.

Due to technological advancements, the transient sky can be explored on ever shorter timescales, resulting in new and unknown astronomical events being found across the entire electromagnetic spectrum and beyond. Reporting and keeping track of the discoveries of all these events is a growing challenge, one that the IVOA has approached with the introduction of VOEvents in 2006. In their Sky Event Reporting Metadata Recommendation\cite{VOEvent}, they describe a standard format to structure metadata about any time-varying astronomical event using \texttt{xml}. The data packet that contains this structured metadata is referred to as a \emph{VOEvent} and the means to transport it to other facilities potentially interested in performing follow-up observations upon the reception of such an event is called the \emph{VOEvent network}. The VOEvent semantics are intentionally kept broad, allowing the VOEvent network to be used for a wide range of applications. Consequently, VOEvents of a certain class are only interpretable by those receivers with common expectations about the contents of the event. Users, those who send events (\emph{authors}) and receive them (\emph{subscribers}), have to agree on what information can be contained within a VOEvent, since only the way of structuring this information is specified by the IVOA recommendation. This paper aims to tailor the use of VOEvents to the needs of the FRB community.

Key information about the VOEvents themselves is given in the IVOA identifiers. Every VOEvent must have a unique and valid International Virtual Observatory Resource Name (IVORN) in order to distinguish between events and to enable subscribers to find specific events or streams of events. An IVORN consists of an author ID, stream ID and a local ID unique to the event (see next section). VOEvent IVORNs are one of the means that subscribers can use to filter messages of interest from the many messages sent over the network. Events can also be filtered based on author, instrument, date, etc. through an \texttt{xml} parser, although specifically scripted VOEvent parsers, such as \texttt{voeventparse}\cite{VOEventParse} are often used. Through the IVORNs it is also possible to find who is responsible for sending a specific event or to create VOEvent archives in which specific events can be queried using the event's IVORN.

\begin{figure}
\vspace{-23pt}
\centering
\includegraphics[width=12cm]{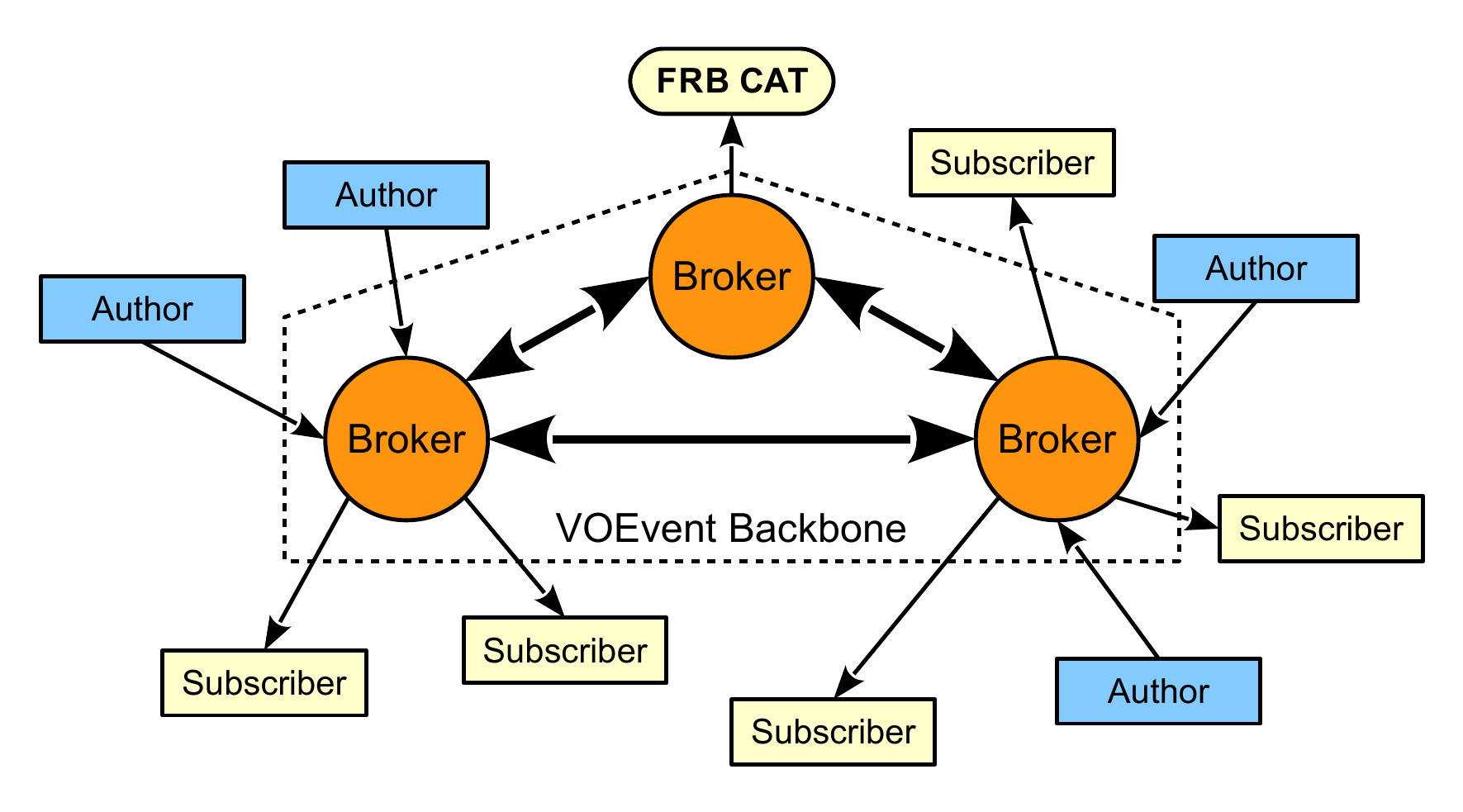}
\vspace{-10pt}
\caption{The VOEvent framework consists of connections between individual Brokers to which events can be pushed by Authors and received by Subscribers. The FRB Catalogue (FRBCAT) will automatically ingest events received by its local broker. More technical details about the connections within the VOEvent Network and how to set them up can be found in the VOEvent Transport Protocol Recommandation\cite{VOEventTransport} and the online documentation provided on \href{https://github.com/ebpetroff/FRB_VOEvent}{GitHub}.\label{fig:VOEvent_network}}
%\vspace{-10pt}
\end{figure}

The VOEvent network is designed according to the VOEvent Transport Protocol\cite{VOEventTransport} and uses Transmission Control Protocol (TCP) connections to transfer a VOEvent error-free from an author to a subscriber. However, to ensure that the network is not flooded with identical events the TCP connections are relayed through a broker (a daemon tool maintained by an institute registered with the IVOA). A broker only re-transmits events to its registered subscribers if the event's body is bit-for-bit different from all the events sent over the network in the last 30 days. The network is thus built upon several interconnected brokers (the VOEvent Backbone) that have registered subscribers connected to one of them and registered authors who are allowed to send VOEvents over the network, shown in Figure~\ref{fig:VOEvent_network}. The author-to-broker connections are temporal connections whereas the connections between brokers and subscribers are always open. Tools, like the software package Comet~\cite{Comet}, have already been developed that can act as brokers or subscribers and take care of the technical requirements to keep the TCP connections alive. These tools make the VOEvent network reliable and easy to use, ideal for rapid transmission of events within a large collaboration or community.

No standard has currently been developed to register a new subscriber. If a facility or individual is interested in receiving VOEvents from a specific author or stream, the first step is to set up a local machine running Comet, or some similar tool, as a subscriber. If the broker of interest requires users to register their IP address, the next step is to get the facility machine white-listed at the broker and implement the necessary filters to only receive events of interest. The same holds true for the registration of a new author. It is to be expected that this method of registration will change in the future when the network has progressed to later stages of its development. The authors will actively participate in this development and provide new participants with help in how to get started on the short term (see section \emph{Code Repository}).

\section*{The FRB VOEvent}
The VOEvent standard defines an \texttt{xml} schema describing how an author should report on an astronomical event but not what should be reported. It is left to the author to decide the content and meaning of the event. If VOEvents are going to be used to report FRBs in the near future, a consensus should be reached about their content beforehand in order to facilitate interpretation by potential subscribers. Therefore we suggest a common usage of VOEvent elements when reporting new bursts or improved FRB parameters. The usage may change if alternative designs are deemed to better accommodate the needs of the community. However, here we present the basic elements and content of an FRB VOEvent which have been agreed upon by a working group in the FRB community. The tags given in the in-line boxes serve as an example on how the required elements of a "Detection FRB VOEvent" are structured. \\

Each FRB VOEvent is contained in a single \texttt{xml} \textbf{$<$VOEvent$>$} element containing several of the sub-elements below.
\begin{lstlisting}
<voe:VOEvent xmlns:xsi="http://www.w3.org/2001/XMLSchema-instance" 
             xmlns:voe="http://www.ivoa.net/xml/VOEvent/v2.0"
             xsi:schemaLocation="http://www.ivoa.net/xml/VOEvent/v2.0 http://www.ivoa.net/xml/VOEvent/VOEvent-v2.0.xsd" 
             version="2.0" role="observation" 
             ivorn="ivo://[institute]/[instrument]#FRB[YYMMDDhhmm]/[MJD]">

    [Main body]

</voe:VOEvent>
\end{lstlisting}
\begin{description}[noitemsep,leftmargin=!,labelwidth=\widthof{\bfseries $<$What$>$},align=right]
   \item[$<$Who$>$] The $<$Who$>$ element provides a subscriber with information about the author. Each facility should have a unique Author IVORN which refers to the institute issuing the event together with the date and time at which the event was created. This IVORN is typically in a reversed DNS format (i.e. \texttt{ivo://de.mpg.mpifr-bonn/contact}). This element should also contain the contact details of the person responsible for the event within the issuing institute, and is a required element in any VOEvent.
\begin{lstlisting}
<Who>
    <AuthorIVORN>ivo://[institute]/contact</AuthorIVORN>
    <Date>[YYYY-MM-DDThh:mm:ss]</Date> <!-- Time of event creation -->
    <Author><contactEmail>[E-mail]</contactEmail><contactName>[Name]</contactName></Author>
</Who>
\end{lstlisting}
   \item[$<$What$>$] What has been observed by the author. This element holds all the scientific parameters describing the observed FRB. Only parameters populated with values should be included in the event. Sample FRB VOEvents and a table describing all parameters are published in the \href{https://github.com/ebpetroff/FRB_VOEvent}{FRB VOEvent GitHub repository} to show what parameters the authors believe can fully describe an FRB, but the community is encouraged to give feedback about the completeness of these lists. Because the intended use of FRB VOEvents can trigger different responses at observatories the scientific parameters are grouped under the following four headers.
   \begin{description}[style=nextline,noitemsep]
      \item[observatory parameters] All the relevant telescope and back-end parameters, like beam size, centre frequency and telescope gain.
\begin{lstlisting}
<Group name="observatory parameters">
    <Param dataType="float" name="beam_semi-major_axis" ucd="instr.beam;pos.errorEllipse;phys.angSize.smajAxis" unit="MM" value=""/>
    <Param dataType="float" name="beam_semi-minor_axis" ucd="instr.beam;pos.errorEllipse;phys.angSize.sminAxis" unit="MM" value=""/>
    <Param dataType="float" name="beam_rotation_angle" ucd="instr.beam;pos.errorEllipse;instr.offset" unit="Degrees" value=""/>
    <Param dataType="float" name="sampling_time" ucd="time.resolution" unit="ms" value=""/>
    <Param dataType="float" name="bandwidth" ucd="instr.bandwidth" unit="MHz" value=""/>
    <Param dataType="float" name="nchan" ucd="meta.number;em.freq;em.bin" unit="None" value=""/>
    <Param dataType="float" name="centre_frequency" ucd="em.freq;instr" unit="MHz" value=""/>
    <Param dataType="int" name="npol" unit="None" value=""/>
    <Param dataType="int" name="bits_per_sample" unit="None" value=""/>
    <Param dataType="float" name="gain" unit="K/Jy" value=""/>
    <Param dataType="float" name="tsys" ucd="phot.antennaTemp" unit="K" value=""/>
    <Param name="backend" value=""/>
    <Param name="beam" value=""><Description>Detection beam number if backend is a multi beam receiver</Description></Param>
</Group>
\end{lstlisting}
      \item[observation parameters] Information about observations currently underway or scheduled to be performed (see next section), including information such as when a dedicated FRB observation started and how long it will last.
      \item[event parameters] Only the essential, easy to extract FRB parameters, such as the DM, width and signal-to-noise ratio of the detected burst. These parameters should allow for rapid follow-up observations with different instruments within seconds to minutes after the initial detection. This element contains the values considered to be the minimum needed to describe an FRB and all parameters in this element should be populated.
\begin{lstlisting}
<Group name="event parameters">
    <Param dataType="float" name="dm" ucd="phys.dispMeasure" unit="pc/cm^3" value=""/>
    <Param dataType="float" name="dm_error" ucd="stat.error;phys.dispMeasure" unit="pc/cm^3" value=""/>
    <Param dataType="float" name="width" ucd="time.duration;src.var.pulse" unit="ms" value=""/>
    <Param dataType="float" name="snr" ucd="stat.snr" value=""/>
    <Param dataType="float" name="flux" ucd="phot.flux" unit="Jy" value=""/>
    <Param dataType="float" name="gl" ucd="pos.galactic.lon" unit="Degrees" value=""/>
    <Param dataType="float" name="gb" ucd="pos.galactic.lat" unit="Degrees" value=""/>
</Group>
\end{lstlisting}
      \item[advanced parameters] Newly obtained or more precisely fit parameters, once the raw FRB data have gone through rigorous analysis. If such parameters are known at the time of issuing the initial detection, they should also be included. The parameters in the latest FRB VOEvent from the detection event's author are taken to be the FRB's true values.
   \end{description}
   \item[$<$WhereWhen$>$] The IVOA Space-Time Coordinate (STC) specifications\cite{STC} are used for this element to describe where on the celestial sky the event has happened and when.
\begin{lstlisting}
<WhereWhen>
    <ObsDataLocation>
        <ObservatoryLocation id="[Instrument location]"/>
        <ObservationLocation>
    	    <AstroCoordSystem id="UTC-FK5-GEO"/><AstroCoords coord_system_id="UTC-FK5-GEO">
    	    <Time unit="s"><TimeInstant><ISOTime>[YYYY-MM-DDThh:mm:ss.ssssss]</ISOTime></TimeInstant></Time> <!-- Time FRB occured -->
    	    <Position2D unit="deg"><Name1>RA</Name1><Name2>Dec</Name2><Value2><C1>[RA in degrees]</C1><C2>[DEC in degrees]</C2></Value2><Error2Radius>[Position error in degrees]</Error2Radius></Position2D>
    	    </AstroCoords>
    	</ObservationLocation>
    </ObsDataLocation>
</WhereWhen>
\end{lstlisting}
   \item[$<$Why$>$] This element contains an importance parameter between zero and one indicating how important the author considers the event to be for follow-up by a subscriber with zero being least important and one being most important for immediate follow-up. The importance element can also be considered as an assurance parameter. This element also supports several sub-elements including a description element that must either be populated with some text describing the score or with a link to a survey webpage where it is described in more detail.
\begin{lstlisting}
<Why importance="0 - 1">
    <Concept>[Flag that importance corresponds to]</Concept>
    <Description>[Elaboration on flag criteria]</Description>
    <Name>FRBYYMMDD</Name> <!-- Name of the FRB -->
</Why>
\end{lstlisting}
   \item[$<$How$>$] Optional element to provide a subscriber with more information on how the detection was made by providing a URL to a website with the back-end or system configuration or survey description, for instance. This element can also contain a link to the data used to create the event if the author made this data publicly available.
   \item[$<$Citations$>$] Optional element, in the case of a new detection but required when issuing an event that relates to a previous FRB VOEvent or VOEvent from a different project. In this case the author has to indicate the event's relation to the previously sent VOEvent and provide the Event IVORN of the VOEvent to which it refers. Allowed event relations are \emph{follow-up} and \emph{supersedes} or \emph{retraction} to refer to FRB VOEvents previously sent by the same author.
\end{description}

These sub-elements provide a subscriber with all the necessary scientific information to decide how to act upon an event, either by taking follow-up observations, start shadowing a telescope, updating the FRBCAT, or waiting for the next VOEvent. The main $<$VOEvent$>$ element is the place holder for information about the VOEvent itself, needed to properly parse the event. For instance, it specifies the version of the VOEvent standard used to structure the metadata in the event. Furthermore, this element contains the ``role'' attribute that depicts the purpose of the event. If an author has a detection to which they want subscribers to respond, the FRB VOEvent should be sent with the ``role'' attribute \emph{observation}. The ``role'' attribute can also be set to \emph{test} if the FRB VOEvent was sent for testing purposes and no action is required by subscribers. The ``role'' \emph{utility} is reserved for events whose main purpose is to report on dedicated FRB observations (see next section). These messages can be used to coordinate observing campaigns that are a cooperation between multiple observatories.

\begin{figure}
%\centering
\hspace{-0.9cm}\includegraphics[width=1.1\textwidth]{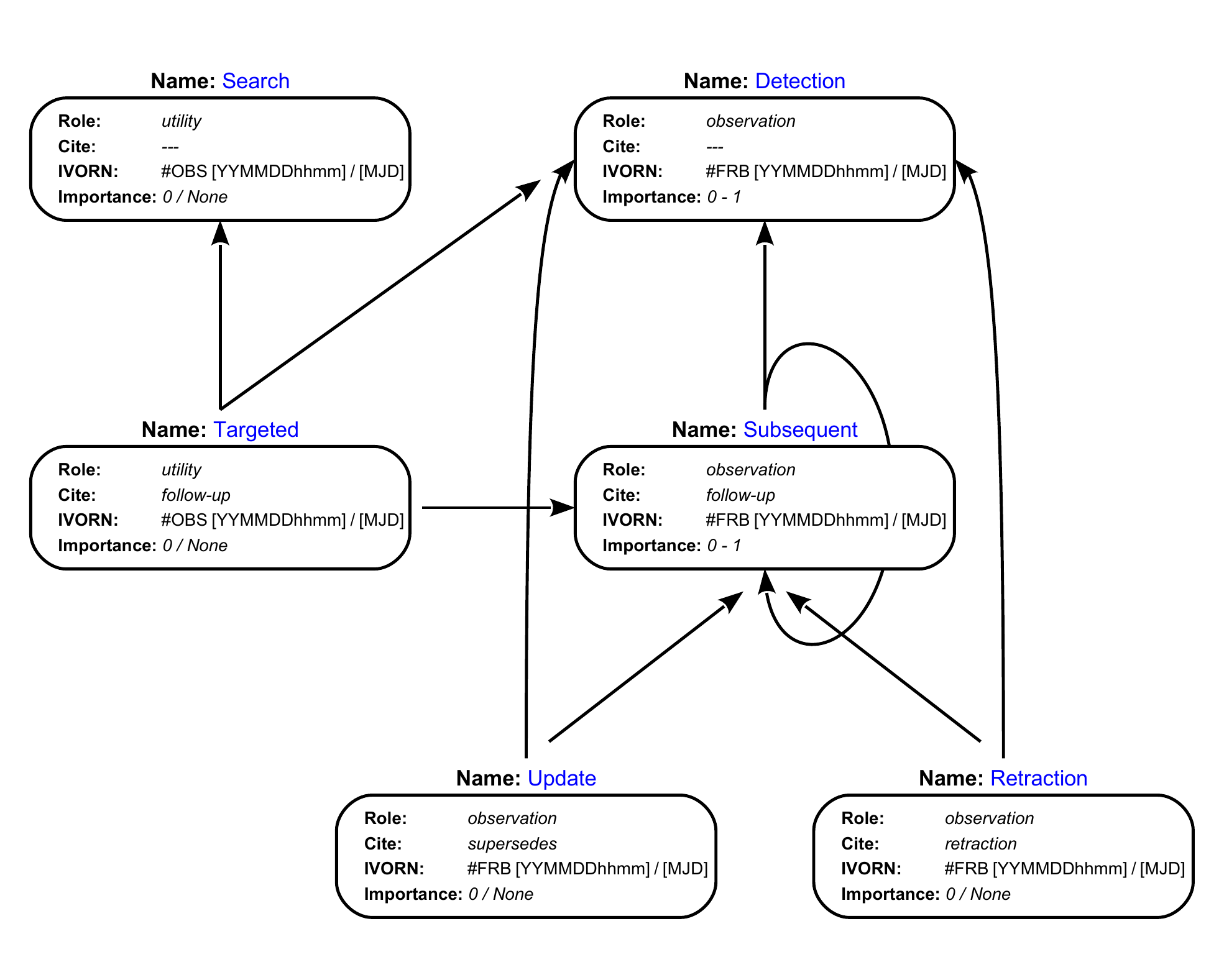}
\caption{The six types of FRB VOEvent with their required VOEvent element parameters. The event type is deduced from these specific parameters and potential citation to another FRB VOEvent. Which type of FRB VOEvent can refer to which other types is therefore limited and indicated here with the black arrows. \label{fig:citerelations}}
\end{figure}

The final important component of the $<$VOEvent$>$ element is the Event IVORN. As discussed above, the Event IVORN is a unique string used to identify a single FRB VOEvent. It should start with \texttt{ivo://} followed by an author ID, stream ID and a local ID with a similar reverse DNS format to the Author IVORN. With a large community potentially interested in sending many FRB VOEvents about the same FRB sources, especially if many are found to repeat, it is sensible to standardise the IVORN allocations for new events. In the case when VOEvents about the same FRB source but from different instruments are broadcast over the network and users would like to organise this information per FRB rather then per instrument this becomes particularly useful. We therefore propose to use the name of the institute issuing the FRB VOEvent as the author ID and the name of the instrument used to detect the FRB as the stream ID. Further, in the future there might be multiple FRB detections a day, so in order to distinguish VOEvents from these events we use the FRB name and time of detection as the local ID in combination with the VOEvent creation MJD. The full Event IVORN will then take the form \texttt{ivo://[institute]/[instrument]\#FRB[YYMMDDhhmm]/[MJD]}. For example, a new detection at 12:00 on 29 March, 2018 using the Apertif instrument by ASTRON would have the IVORN \texttt{ivo://nl.astron/apertif\#FRB1803291200/58206.50000000}. This enables the community to publish FRB VOEvents under the same FRB name, but gives a natural way to sequence the events. If an event reports on an observation rather then the detection of an FRB we recommend to use the abbreviation ``OBS'' instead of ``FRB'' in the local ID. Examples of the use of these Event IVORNs are given in Figure~\ref{fig:citerelations}. A sequential history of VOEvents for a given FRB will be given in the FRB Catalogue, as discussed in a later section.

%\begin{figure}
%\centering
%\includegraphics[width=14cm]{Figures/DiffVOEvent1.pdf}
%\caption{An example of a series of VOEvents issued concerning a particular FRB source. The source is first detected with a radio telescope and an \emph{observation} event is issued with IVORN index 0000. If the source is observed with other telescopes they can report their findings using the \emph{follow-up} keyword with a citation to the original detection and an IVORN index incremented by one (0001 and 0002). If detection facility makes further fits to the FRB and reports updated parameters to FRBCAT they use a \emph{utility} VOEvent citing the original detection with the keyword \emph{supersedes} and the next IVORN index in the series (0003). \label{fig:VOEvent}}
%\end{figure}

\section*{FRB VOEvent Types}
Since FRB VOEvents are envisioned to trigger a variety of responses at observatories, their content should be adapted to the purpose of the event. Currently, six unique message types have been identified to allow communication about new detections, follow-up observations or the start of dedicated FRB observation campaigns. Each of the types is given an event type name, appointed a selection of required VOEvent element parameters and assigned a citation relation shown in Figure~\ref{fig:citerelations}.

Four of the FRB VOEvent types can be used to report information on actual astronomical signals. These types are:
\begin{description}[noitemsep,leftmargin=!,labelwidth=\widthof{Subsequent},align=right]
   \item[Detection] Used to report the \emph{detection} of a \emph{new} FRB. No $<$Citations$>$ element has to be given because it is the first message describing this event.  Required elements include the ``observatory parameters'' and ``event parameters'' in the $<$What$>$ element and an importance value in the $<$Why$>$ element between 0 and 1 to broadcast how relevant the author considers the event to be. Setting the event's ``role'' to \emph{observation} tells any subscriber that the FRB VOEvent reports on an astronomical event.\newpage
   \item[Subsequent] Any FRB VOEvent that describes another detection signal of a previously reported event, for instance the same event in a different part of the electromagnetic spectrum or a repeat burst, is considered a \emph{subsequent} event. This type of event must therefore reference events of the same type or the type \emph{detection} and a $<$Citations$>$ element must now be given. The rest of this types content is similar to the content of a \emph{detection} FRB VOEvent. Note that follow-up observations without a detection can be reported using a \emph{targeted} FRB VOEvent, see below.
   \item[Update] Once a more rigorous analysis of the data has been performed and more information about the event is available, an \emph{update} can be sent to link the new parameters with the old. \emph{Updates} can also update event parameters that are already reported either in a \emph{detection} or \emph{subsequent} FRB VOEvent. A $<$Citations$>$ element is required as well as the additional ``advanced parameters'' group in the $<$What$>$ element that contains the newly obtained event parameters. If the intended purpose of the event is purely to update the catalogue, the importance flag in the $<$Why$>$ element should be set to 0 to prevent observatories from accidentally confusing this event type with a \emph{detection} and start performing follow-up observations upon the reception of an \emph{update} FRB VOEvent.
   \item[Retraction] In the case where rigorous (offline) analysis reveals a previously reported event to be non-astrophysical, for instance to be radio frequency interference (RFI), then a \emph{retraction} FRB VOEvent can be used to retract all previously sent messages about this event. With the transmission of a \emph{retraction} message one thus broadcasts that any planned or ongoing efforts to obtain more data on the event referred to by this message should be terminated. A \emph{retraction} FRB VOEvent will also trigger the removal of its FRB Catalogue entry (see next section). The most important parameters in a \emph{retraction} FRB VOEvent are therefore the $<$Citations$>$ element and the importance flag in the $<$Why$>$ element. The former to ensure that the correct FRB VOEvent is retracted and the later to prevent any new observations from being performed on the reported event.
\end{description}

Discussions during the Aspen 2017 Winter Conference on Fast Radio Bursts revealed an interest to not only be able to share detection metadata with the community, but also to share details on planned dedicated FRB observations. The main motivation for this extra functionality is the lack of detections in recently performed follow-up observations triggered by a real-time FRB detection at Parkes~\cite{PetroffFRB,Keane2016,131104Radio,FRB150215}. Follow-up observations might always be performed too late if prompt emission or afterglows at other frequencies are very short-lived, making commensal observations a more attractive option for a multi-wavelength measurement. This could also enable an event occurring at other wavelengths to be linked with an FRB seen in the radio regime. If observatories share their scheduling of dedicated FRB observations, others can decide to start shadowing one of these telescopes to increase the chance of making a multi-wavelength detection of an FRB. Two extra FRB VOEvent types are proposed here for this purpose:
\begin{description}[noitemsep,leftmargin=!,labelwidth=\widthof{Subsequent},align=right]
   \item[Search] The type \emph{search} is used to report the start of a blind search, an observation with no particular assigned target. Currently these types of observations are performed in surveys by radio observatories trying to discover new pulsars, which run a single pulse search engine in parallel to discover FRBs. The content of the \emph{search} message is therefore biased towards radio telescopes, but the parameters included in the ``observatory parameters'' and ``observation parameters'' list, required for this type of event, can be expanded when deemed necessary. Aside from these two groups in the $<$What$>$ element, the position information in the $<$WhereWhen$>$ element must be set to the boresight coordinates of the telescope or beamformed area of the array and to the date and time it will be pointed there. The radius of the full field of view (FoV) of a telescope or beamformed array is specified as the pointing position error. In the case where multiple sky positions are observed simultaneously by multiple beamformed areas or telescopes, a seperate FRB VOEvent should be created for every single sky position. Since these parameters do not refer to the location of a transient astrophysical event, the VOEvent ``role'' should be set to \emph{utility} and the importance flag in the $<$Why$>$ element must be 0 or None. The message does not necessarily need to be followed-up by other telescopes. To stress the non-astrophysical nature of the event it is recommended to change the event's IVORN local ID from \texttt{\#FRB[YYMMDDhhmm]/[MJD]} to \texttt{\#OBS[YYMMDDhhmm]/[MJD]}.
   \item[Targeted] Whenever observatories decide to start shadowing another telescope, re-observe known FRB locations or try to find a connection with other astrophysical phenomena they can report the start of their efforts with a \emph{targeted} FRB VOEvent. This type is similar to a \emph{search} VOEvent with the only exception that it must contain the $<$Citations$>$ element, referring to a \emph{targeted} or \emph{search} FRB VOEvent to specify which telescope is being shadowed. Additionally, a \emph{targeted} event can cite a \emph{detection} or \emph{subsequent} FRB VOEvent when an observatory re-observes an FRB location in order to find repeat bursts. The event can also cite a VOEvent from another project if a source is observed in an effort to associate it to FRBs.
\end{description}
A \emph{search} FRB VOEvent can be considered to be the counterpart of a \emph{detection} message and so is the \emph{targeted} event type the counterpart of the \emph{subsequent} type, only they report on a planned or ongoing observation rather then a detection of an FRB. These two types can therefore be used to coordinate follow-up efforts on the transmission of a \emph{detection} FRB VOEvent as described in the third use case from the motivation section.
\newpage

\section*{Connection with the FRB Catalogue}
The FRB Catalogue has been upgraded to automatically ingest FRB VOEvents, parse them, and populate the catalogue in real-time as events are issued by authors. This functionality has been developed to provide a simple way to add bursts to the catalogue (until now this has been done manually) and to provide a centralised hub for information about new detections as a service for the community. 

The FRB Catalogue will receive all events transmitted or received by the ASTRON broker -- a Comet-based broker service managed and maintained by ASTRON -- and will automatically place these in the catalogue. Upon notification of a new FRB a new page will be created in the catalogue for the source. Additional detections, all with the ``role'' \emph{observation} citing the original FRB detection, will be added to the catalogue as new observations of an existing FRB. If new fits or updated parameters are available for an observed pulse after additional processing the best-fit parameters in the catalogue can be updated using an \emph{update} FRB VOEvent; in this case the event should cite the VOEvent containing the old parameters with the citation keyword \emph{supersedes}. Events can also be removed from the catalogue using a \emph{retraction} FRB VOEvent. The catalogue will keep a VOEvent history for each FRB; however, VOEvents concerning retracted FRBs will not be kept on the catalogue webpage. 

%% Added the following two paragraphs to better describe some things that were unclear
All events received will be added to the catalogue database; though, not all events will be visible on the default homepage. Events that have been verified, either through publication, an \textit{update} VOEvent, or a very high importance value for the \textit{detection} message ($>$0.95) will be displayed by default. All events that have been received (but not retracted) are available on the catalogue webpage by toggling the ``Show all/Show verified'' button in the catalogue header. Retracted messages will only be accessible via the catalogue's database.

With regards to possible race conditions if two telescopes should simultaneously detect the same FRB: the automatic functionality of the catalogue will create two distinct event entries for this burst to be merged by the catalogue administrator. At the time of merging, the event received with the earliest timestamp will be considered the \textit{detection} and all others will be given the designation \textit{subsequent}.

We would like to conclude with a few notes on future-proofing as the FRB population grows and evolves. As more physical properties of FRB sources are understood, new parameters to describe FRBs may become necessary and new naming conventions may arise. The FRB Catalogue is not intended as a static resource but has been developed flexibly with the aim to add or remove parameters and provide functionality desired by the community. Continued feedback from users of the catalogue on the features they would like to see is welcomed and appreciated and can be send to the authors of this paper. It is the intention of the team responsible for this catalogue to implement viable new suggestions as quickly as possible and to adapt the catalogue to best capture the essential information for FRB science.

\section*{Discussion and Conclusions}

In this paper we provide a proposed standard for FRB VOEvents -- flexible, machine-parsable and human-readable \texttt{xml} files used to disseminate astronomical events. VOEvents have been successfully used in the gamma-ray burst and supernova communities to notify facilities around the world of new transient phenomena and to facilitate rapid multi-wavelength follow-up. As new efforts commence to find FRBs with telescopes around the world, a large increase in the population is expected and real-time detections will become commonplace. The VOEvent framework is well suited to sending FRB triggers as the network infrastructure is already in place and is maintained by the International Virtual Observatory Alliance (IVOA). The structure of VOEvents themselves is extremely flexible and the standard contents of VOEvents sent about a certain source class must be defined by the community that intends to send and receive these events.

The VOEvent templates located in the associated GitHub repository are intended to provide the community with a standard that can be used globally for FRB triggers. This standard is intended to be flexible, modular, and useful to the largest group possible to enable high impact, rapid response transient science. The standard presented here has been developed by a subset of the FRB community developing large surveys to detect FRBs, but the ultimate format and structure of FRB VOEvents should be decided on in collaboration with a majority of current users of such a system. As such, we encourage all users who are interested in this initiative to join this working group by contacting the authors of this paper. We also welcome feedback on any aspect of this effort by members of the community as well as suggestions for changes or improvements that could be made to this model.

\newpage
\section*{Code Repository}

Useful tools related to FRB VOEvents including \texttt{xml} templates, example events, and simple scripts are available in a GitHub repository for community use at \url{https://github.com/ebpetroff/FRB\_VOEvent}. The templates for each VOEvent type provide a list of all the parameters available for that type. FRB VOEvents sent through the VOEvent Network do not need to include all the parameters presented in the templates and only populated parameters should be included in the file. The FRB VOEvents in the repository are examples of what real events coming over the network will look like; these can be used for testing local parsers. Publicly available open-source code for parsing VOEvents is also available through \texttt{python} packages\cite{VOEventParse}. The repository also contains documentation for the events such as parameter definitions and information about what parameters and elements are required for each event type. As the network develops, up to date information on how to get started and connected to the FRB VOEvent Framework will be published here too.

\section*{Acknowledgments}

The authors would like to thank J. Swinbank for helpful comments and input as well as the participants of the 2017 Aspen Winter Conference Fast Radio Bursts: New Probes of Fundamental Physics and Cosmology.
Funding from the European Research Council under the European Union's Seventh Framework Programme (FP/2007-2013) / ERC Grant Agreement n. 617199 (EP and JvL).
Funding from the Netherlands eScience Center under grant AA-ALERT, 027.015.G09 (OMR and RH).
Funding from the Research Corporation for Scientific Advancement (RCSA) (LH).
AW is funded by the South African Research Chairs Initiative of the DST and NRF of South Africa

\bibliography{FRB_VOEvents}

\end{document}